\documentclass[]{aastex631}

\usepackage{soul}

\begin{document}

\title{Smallest scale clumpy star formation in Stephan's Quintet revealed from UV and IR imaging}

\author[0000-0003-1409-1903]{P. Joseph}
\affiliation{Indian Institute of Astrophysics\\
Koramangala II Block, Bangalore, India}
\affiliation{Department of Physics\\
Christ University, Bangalore, India}

\author{K. George}
\affiliation{Faculty of Physics \\
Ludwig-Maximilians-Universit{\"a}t\\
Scheinerstr. 1, Munich, 81679, Germany}

%\collaboration{20}{(AAS Journals Data Editors)}

\author{S. Subramanian}
\affiliation{Indian Institute of Astrophysics\\
Koramangala II Block, Bangalore, India}

\author{C. Mondal}
\affiliation{Inter-University Centre for Astronomy and Astrophysics\\
Ganeshkhind, Post Bag 4, Pune 411007, India}

\author{A. Subramaniam}
\affiliation{Indian Institute of Astrophysics\\
Koramangala II Block, Bangalore, India}



\begin{abstract}
The spatial distribution and physical sizes of star forming clumps at the smallest scales provide valuable information on hierarchical star formation (SF). In this context, we report the sites of ongoing SF at $\sim$120 pc along the interacting galaxies in Stephan's Quintet (SQ) compact group using AstroSat-UVIT and JWST data. Since ultraviolet radiation is a direct tracer of recent SF, we identified star forming clumps in this compact group from the FUV imaging which we used to guide us to detect star forming regions on JWST IR images. The FUV imaging reveals star forming regions within which we detect smaller clumps from the higher spatial resolution images of JWST, likely produced by PAH molecules and dust ionised by FUV emission from young massive stars. This analysis reveals the importance of FUV imaging data in identifying star forming regions in the highest spatial resolution IR imaging available. 
%This example manuscript is intended to serve as a tutorial and template for
%authors to use when writing their own AAS Journal articles. The manuscript
%includes a history of \aastex\ and includes figure and table examples to illustrate these features. Information on features not explicitly mentioned in the article can be viewed in the manuscript comments or more extensive online
%documentation. Authors are welcome replace the text, tables, figures, and
%bibliography with their own and submit the resulting manuscript to the AAS
%Journals peer review system.  The first lesson in the tutorial is to remind
%authors that the AAS Journals, the Astrophysical Journal (ApJ), the
%Astrophysical Journal Letters (ApJL), the Astronomical Journal (AJ), and
%the Planetary Science Journal (PSJ) all have a 250 word limit for the 
%abstract\footnote{Abstracts for Research Notes of the American Astronomical 
%Society (RNAAS) are limited to 150 words}.  If you exceed this length the
%Editorial office will ask you to shorten it. This abstract has 161 words.

\end{abstract}



\section{Detection of smallest star forming knots in the IGM} \label{sec:intro}
%\pjnew{I think the title might be long, perhaps it can
%be shortened to "Detection of smallest star-forming knots".}

The spatial distribution of the star forming clumps inherit the spatial
properties of the gas from which they form. They can therefore be used
as tracers to understand the physical processes associated with the star
formation. The distribution of star forming clumps and their physical sizes at the smallest scales provide valuable information on the hierarchical structures involved in star formation. But all these properties are strongly dependent on the properties of the host galaxy environment \citep{menon2021dependence}. Hence study of hierarchical nature of star formation in different galactic environments can provide valuable insights on our understanding of star formation processes. Such previous studies were restricted to very nearby galaxies due to the limited spatial resolution of the telescopes used for the detection of star forming regions at sub kpc scales. Several studies \citep{peeters2004polycyclic, xie2019new, whitcomb2020comparative}
% ({\bf https://ui.adsabs.harvard.edu/abs/2004ApJ...613..986P/abstract,https://ui.adsabs.harvard.edu/abs/2019ApJ...884..136X/abstract,https://ui.adsabs.harvard.edu/abs/2020ApJ...901...47W/abstract}) 
have shown that emission features in 6.2, 7.7, 8.6, and 11.3 {\textmu} (mid-IR) from the excited PAH molecules can be used to trace the location of FUV-bright young stars (which excites the PAH molecules) and hence recent star formation. The superior spatial resolution JWST IR images of SQ is the best available data to study SF  at the smallest scales in the intragroup medium (IGM). So with the superior spatial resolution IR observations from JWST, we can now trace the spatial distribution of star forming clumps and their physical sizes in galaxies at farther distances. However, the Spitzer/IRS observations have shown that PAHs can be destroyed by intense starbursts and are also absent in low metallicity galaxies \citep{li2020spitzer, o2006polycyclic}.
% ({\bf add references, https://arxiv.org/pdf/2003.10489.pdf, https://iopscience.iop.org/article/10.1086/500529/pdf}). 
Again, PAH emission does not necessarily trace exclusively young stars as PAHs can also be excited by visible photons \citep{li2002infrared}.
% ({\bf add references, https://arxiv.org/pdf/astro-ph/0201060.pdf}). 
Hence it is essential to use a direct star formation (SF) tracer and effectively use that to guide us to detect star forming regions on JWST IR images.

Ultraviolet light produced by young OBA stars is perhaps the only direct tracer of ongoing star formation ($<$ 10${^8}$ yrs) with not much contamination. In this context, we study the spatial distribution and physical sizes of star forming clumps in the IGM of a nearby compact group of galaxies, Stephan's Quintet,  using a combined analysis of AstroSat-UVIT and JWST \citep{pontoppidan2022jwst} images. We demonstrate the power of UV in showing the star forming regions on the JWST IR images in the color composite Fig. 1a and 1b as traced in blue color.\footnote{The JWST data used in this paper can be found in MAST: \dataset[10.17909/qj5k-mm71]{http://dx.doi.org/10.17909/qj5k-mm71}.}(also hosted at \url{https://prajwel.github.io/stephans_quintet/}). We detected smaller scale IR clumps from the JWST image within the FUV emitting regions in SQ, which confirms the recent SF nature of the IR clumps. 
%The star forming segments we detected from FUV image contains smaller knots in IR imaging.
%The F770W image of SQ along with the FUV and MIR segments is shown in Fig 4. 
F770W imaging of these clumps traces the 7.7{\textmu} emission coming from PAH molecules excited by FUV photons. This  highlights the clumpy nature of dust clouds associated with the star-forming regions that we are probing. 
%The 35 sources detected from FUV imaging is matched with 78 sources detected from JWST F770W imaging and is marked in Table 1.
These are regions where gas is collapsing outside the galaxies forming new stars. The size distribution of these IR clumps detected inside the segments identified from UV imaging is shown in Fig 1c. Thanks to JWST resolution, the IR clumps are detected of size as small as 120 pc within FUV clumps of larger area (due to the coarser resolution of UVIT). We note that this is similar to the typical size of OB associations (size $<$ 150pc) detected in many nearby galaxies including the Galaxy \citep{mel1995new,bresolin1998hubble,ivanov1996ob,bastian2007hierarchical,mondal2018uvit}. The identified F770W IR sub-clumps within the FUV clumps basically trace the PAH molecules embedded in dust-obscured young star-forming regions. Therefore, the detection of clumps from JWST F770W image of SQ having size as small as $\sim$ 120 pc signifies that the young star-forming regions in SQ have sizes similar to the OB associations (which are most efficient to excite the PAH molecules to emit in 7.7\textmu) detected in nearby galaxies. This further demonstrates the unique capability of JWST to probe star-forming regions up to much smaller scales in IR than what was possible earlier with Spitzer. Since UVIT has $\sim$ 6 times larger PSF than the JWST/MIRI F770W, the UV segments detected in this study should be mostly composed of multiple smaller clumps. The size of UV segments therefore signifies the extent of the large hierarchical structure composed of smaller star-forming clumps (Similar values reported in nearby systems \citealt{grasha2017hierarchical, rodriguez2020hierarchical, mondal2021tracing}).
Detailed follow-up studies will help to establish the connection between the FUV radiation and PAH excitation in star forming regions within the intergalactic medium.
UV imaging with a resolution similar to that of JWST will be able to probe such stellar clumps directly. 
%Until the availability of such UV instruments, JWST MIR imaging will serve the best to disentangle (though indirectly) the nature of young star-forming clumps in distant galaxies.\\
 We note that the future planned UV survey telescopes like INSIST with JWST like resolution when launched will be a great boost in interpreting JWST imaging data for ongoing star formation \citep{subramaniam2022overview}.

% \begin{figure}[ht!]
% \plotone{F148W_F770W_histogram.png}
% \caption{Distribution of the size of segments (diameter in arcsec and pc)
% detected from UVIT F148W imaging data (blue) and the corresponding segments contained within FUV segments detected from JWST F770W imaging
% data (orange). 
% \label{fig:general}}
% \end{figure}

\begin{figure}
\centering
\includegraphics[width=0.48\textwidth]{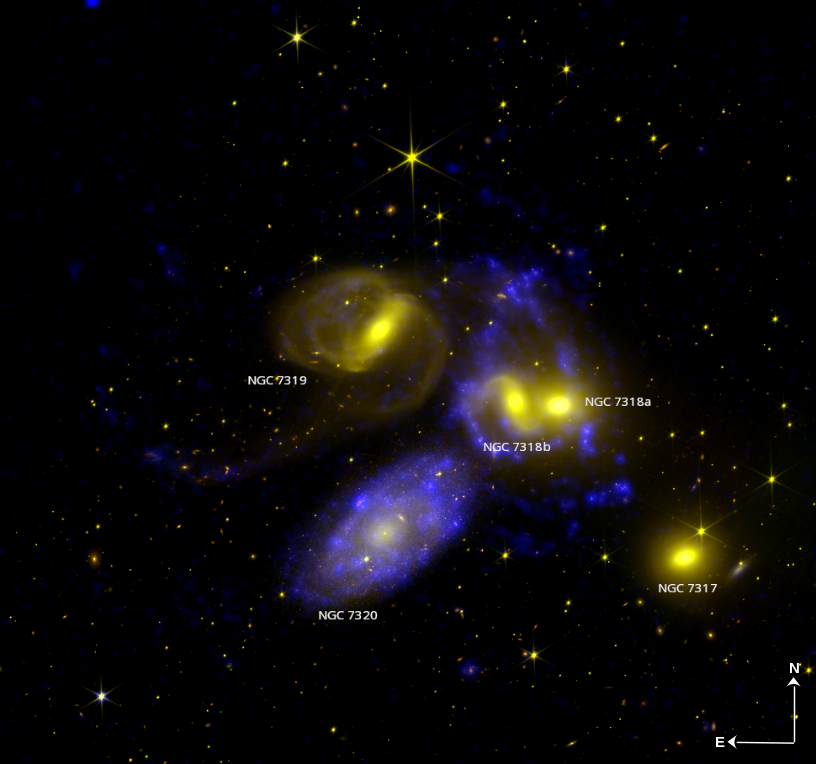}\hfill
\includegraphics[width=0.48\textwidth]{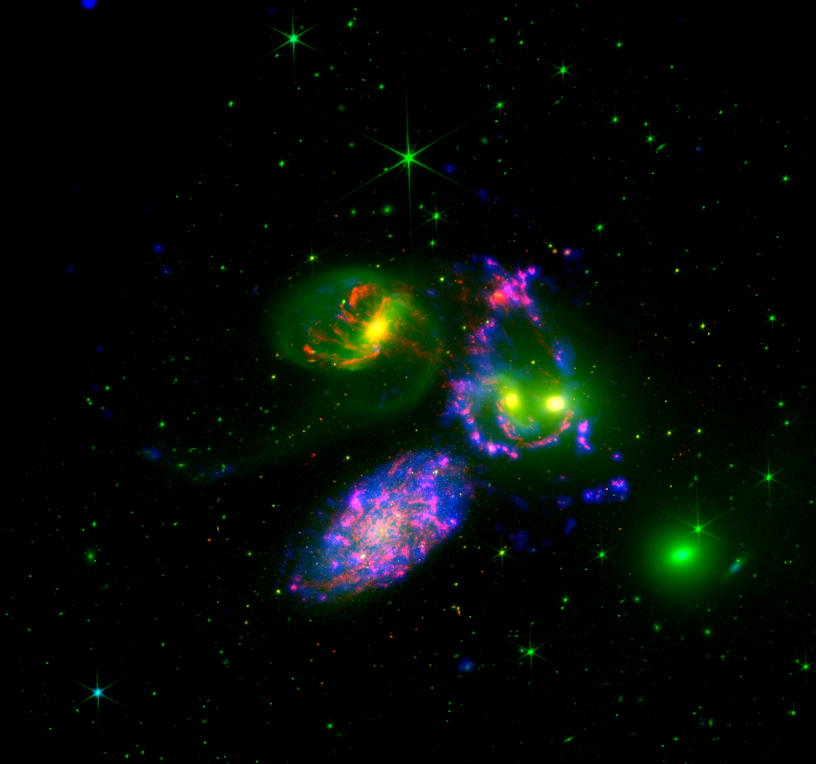}\hfill
\includegraphics[width=0.48\textwidth]{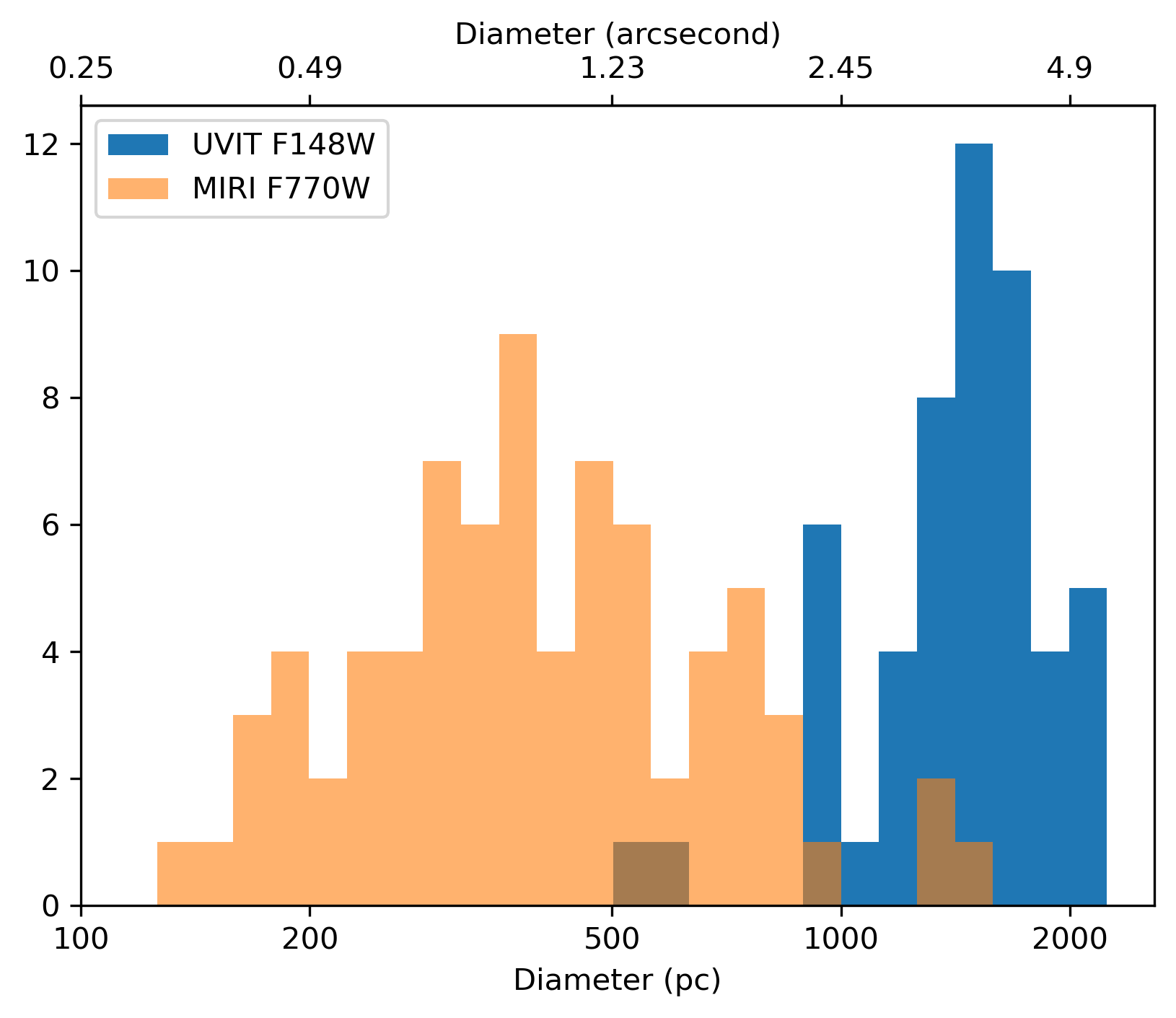}
\caption{\textit{Clockwise from the top-left panel:} a) RGB image of SQ created by assigning red (JWST F200W band), green (JWST F150W band) and blue (UVIT F148W+F169M band) colours to the filter passband images. b) RGB image of SQ created by assigning red (JWST F770W band), green (JWST F200W band) and blue (UVIT F148W+F169M band) colours to the filter passband images. c) Distribution of the size of segments (diameter in arcsec and pc).}
\label{fig:figure16}
\end{figure}

% \usepackage{caption}
% \usepackage{subcaption}

% \begin{figure}
%      \centering
%      \begin{subfigure}[b]{0.3\textwidth}
%          \centering
%          \includegraphics[width=\textwidth]{sources_marked.png}
%          \caption{test}
%          \label{fig:y equals x}
%      \end{subfigure}
%      \hfill
%      \begin{subfigure}[b]{0.3\textwidth}
%          \centering
%          \includegraphics[width=\textwidth]{right_panel.png}
%          \caption{test}
%          \label{fig:three sin x}
%      \end{subfigure}
%      \hfill
%      \begin{subfigure}[b]{0.3\textwidth}
%          \centering
%          \includegraphics[width=\textwidth]{F148W_F770W_histogram.png}
%          \caption{test}
%          \label{fig:five over x}
%      \end{subfigure}
%         \caption{Three simple graphs}
%         \label{fig:three graphs}
% \end{figure}

\begin{acknowledgments}
SS acknowledges support from the Science and Engineering Research Board, India through a Ramanujan Fellowship. AS acknowledges support from SERB Power Fellowship. This research made use of Astropy, a community-developed core Python package for Astronomy \citep{robitaille2013astropy, price2018astropy}. This publication uses the data from the AstroSat mission of the Indian Space Research  Organisation  (ISRO),  archived  at  the  Indian  Space  Science  Data Centre (ISSDC).  UVIT  project  is  a  result  of collaboration  between  IIA,  Bengaluru,  IUCAA,  Pune, TIFR, Mumbai, several centres of ISRO, and CSA. This work is based in part on observations made with the NASA/ESA/CSA James Webb Space Telescope. The data were obtained from the Mikulski Archive for Space Telescopes at the Space Telescope Science Institute, which is operated by the Association of Universities for Research in Astronomy, Inc., under NASA contract NAS 5-03127 for JWST. 
\end{acknowledgments}

%\bibliography{sample631}{}
\bibliography{paper}{}
\bibliographystyle{aasjournal}



\end{document}